\def\twocolumnmode{1}
\if\twocolumnmode0
\documentclass[12pt,a4paper,draftclsnofoot,onecolumn]{IEEEtran}
\else
\documentclass[10pt,a4paper]{IEEEtran}
\fi
\newcommand{\ms}[1]{\if\twocolumnmode0\dimexpr#1/2\relax\else#1\fi}

\usepackage{bm}
\usepackage{cite}
\usepackage{graphicx}
\usepackage{orcidlink}
\usepackage{acronym}
\usepackage{optidef}
\usepackage{upgreek}
\usepackage{tikz}
\usepackage{tikzscale}
\usepackage{pgfplots}
\usepackage{microtype}
\usepackage{makecell}
\usepackage[dvipsnames]{xcolor}
\usepackage[all=normal,mathspacing=tight,paragraphs=tight]{savetrees}

\usetikzlibrary{calc}
\usetikzlibrary{external}
\acrodef{PD}{photodetector}
\acrodef{OWC}{optical wireless communication}
\acrodef{IM/DD}{intensity modulation and direct detection}
\acrodef{RF}{radio-frequency}
\acrodef{SIMO}{single-input multiple-output}
\acrodef{FSO}{free-space optical}
\acrodef{UOWC}{underwater optical wireless communication}
\acrodef{VLC}{visible-light communication}
\acrodef{SNR}{signal-to-noise ratio}
\acrodef{BER}{bit error rate}
\acrodef{TIA}{transimpedance amplifier}
\acrodef{LG}{Laguerre-Gaussian}
\acrodef{LiFi}{light fidelity}

\title{\huge Performance Scaling Laws for PD Array-based Receivers in IM/DD Optical Wireless Communication Systems}
\author{
Aravindh~Krishnamoorthy\textsuperscript{\orcidlink{0000-0001-7186-121X}}\IEEEauthorrefmark{1},~\IEEEmembership{Member,~IEEE,} %
Robert~Schober\IEEEauthorrefmark{2},~\IEEEmembership{Fellow,~IEEE,} %
Harald~Haas\IEEEauthorrefmark{1},~\IEEEmembership{Fellow,~IEEE} \\
{%
	\bigskip
	\small
	\IEEEauthorrefmark{1}~LiFi Research and Development Centre, Electrical Engineering Division, University of Cambridge, UK,\\
	\IEEEauthorrefmark{2}~Institute for Digital Communications, Friedrich-Alexander-Universität Erlangen-Nürnberg, Germany\\
}\vspace{-1cm}%
}
\begin{document}
\bstctlcite{IEEE:BSTcontrol}
\maketitle
\begin{abstract}
We study the performance scaling laws for electrical-domain combining in photodetector (PD) array-based receivers employing intensity modulation and direct detection, taking into account the inherent square-law relationship between the optical and electrical received powers. The performance of PD array-based systems is compared, in terms of signal-to-noise ratio (SNR) and achievable rate, to that of a reference receiver employing a single PD. Analytical and numerical results show that PD arrays provide performance gains for sufficiently narrow beams and above an SNR threshold. Furthermore, increasing the number of PDs alone does not enhance performance, and joint optimization of beam pattern, transverse electromagnetic mode, received power, and PD positions is necessary. Our model and derived insights provide practical guidelines and highlight the trade-offs for the design of next-generation high-bandwidth PD array receivers.
\end{abstract}
\section{Introduction}
Laser-based \ac{IM/DD} \ac{OWC} systems find widespread use owing to their superior performance and low complexity. In these systems, at the transmitter, a laser's output intensity is modulated by an electrical signal, and at the receiver, the photocurrent from the received optical signal is directly converted to an electrical signal via a \ac{TIA}. Hence, unlike coherent systems \cite{GuptaCoherent2019}, no output phase control and local oscillators are needed at the transmitter and receiver, respectively, which lowers the system complexity significantly.

However, since quadrature modulation cannot be utilized in \ac{IM/DD} systems, their spectral efficiency is half that of coherent optical systems \cite{Lapidoth2009,Chaaban2020}. Furthermore, their sensitivity is low due to the lack of local oscillators. Nevertheless, with careful beam management and by exploiting the large, unlicensed bandwidth at optical frequencies, \ac{IM/DD} links can exceed data rates of $1$ Tbps \cite{Fernandes2022,St-Arnault2025}.

The capacity and achievable rates of \ac{IM/DD} systems have been studied in detail \cite{Lapidoth2009,Chaaban2016,Chaaban2020}. Furthermore, technologies such as \ac{LiFi} \cite{Haas2019} based on \ac{IM/DD} systems have been proposed. More recently, in order to enhance performance, robustness to misalignment, and design and deployment flexibility, \ac{PD} array-based receivers have been proposed \cite{Zeng2017Imaging,Koonen2020,Madhavan2020PDArray,Umezawa2021,Sarbazi2024,Jiang2025}.

In \ac{PD} array-based receivers, instead of a single \ac{PD}, an array of \acp{PD} is utilized. Furthermore, the electrical signals generated by the incident beam are optimally combined in the electrical domain. Moreover, since the active areas of the \acp{PD} are smaller compared to a single reference \ac{PD}, their bandwidth can be enhanced via careful design \cite{Alexander1997}. However, in square-law detection systems, such as \ac{IM/DD} systems, the received electrical power is proportional to the square of the received optical power \cite{Alexander1997,Kahn1997}. Taking this into account is important for systems that combine signals in the electrical domain, see \cite{Agrawal2010,Bashir2020}. However, a systematic study of the impact of the square-law relationship is not available in the literature.

Therefore, in this letter, we study the performance scaling laws for electrical-domain combining in \ac{PD} arrays, taking into account the above square-law relationship. In particular, we compare their performance to a single reference \ac{PD} for  Gaussian, \ac{LG}, and uniform laser beam patterns, which realize different levels of beam concentration on the \ac{PD} array. We consider two different design choices: (1) The combined active area of the \ac{PD} array scales with the number of \acp{PD}, (2) the combined active area remains constant, i.e., the active areas of the individual \acp{PD} shrink. We provide intuition to guide the design of next-generation \ac{PD} array-based systems. The contributions of this letter are as follows.
\begin{itemize}
	\item We develop a model for the \ac{SNR} and achievable rates for \ac{PD} array-based receivers, taking into account the square-law relationship between the optical and electrical powers as well as the impact of the \ac{PD} size on bandwidth.
	\item We introduce a \emph{loss factor} and use it to model the extent of suboptimality for different beam patterns, and to compare the performance of \ac{PD} array-based receivers for the two considered design choices with that of a reference receiver employing a single \ac{PD}.
\end{itemize}
Through analysis and numerical simulations, we show that \ac{PD} arrays provide a performance gain when the incident beam is sufficiently narrow and when the received \ac{SNR} is above a threshold, which depends on the incident beam pattern. Furthermore, we show that, in most cases, the performance of the reference receiver can be exceeded by increasing the received power and not by increasing the number of \acp{PD} alone. Lastly, we show that, for \ac{PD} array-based receivers, direct detection of higher-order transverse electromagnetic laser beam modes is suboptimal, and a mode conversion to the fundamental Gaussian beam is beneficial.

The remainder of this paper is organized as follows. In Section \ref{sec:prelim}, we introduce the receiver model and review the factors affecting the bandwidth of \acp{PD}. Furthermore, we describe the considered beam patterns. In Section \ref{sec:combining}, we derive the performance scaling laws. Numerical results and discussions are provided in Section \ref{sec:numerical}, and the paper is concluded in Section \ref{sec:conclusion}.

\section{Preliminaries}
\label{sec:prelim}
In this section, we provide the receiver model, discuss briefly the factors affecting \ac{PD} bandwidth, and describe the considered laser beam patterns.

\subsection{Receiver Model}
\label{sec:model}
We consider a point-to-point \ac{IM/DD} \ac{OWC} system comprising a transmitter and a receiver, and focus on the receiver design. The receiver is composed of a \ac{PD} array with $M$ identical \acp{PD} and employs linear signal processing to combine the received signals from the \acp{PD}, after the \ac{TIA}, in the electrical domain. Let the \emph{optical} power received by the $m$-th \ac{PD} be denoted by $P^{\mathrm{Rx,O}}_m,$ $m=1,\dots,M,$ and let $P^{\mathrm{Rx,O}}_{\mathrm{tot}} = \sum_{m = 1}^{M} P^{\mathrm{Rx,O}}_m$ denote the total received power. The received \emph{electrical} power for the $m$-th \ac{PD} is given by $P^{\mathrm{Rx,E}}_m = (R_{\mathrm{PD}} P^{\mathrm{Rx,O}}_m)^2$ as an inherent consequence of square-law detection \cite{Kahn1997,Hranilovic2005}, where $R_{\mathrm{PD}}$ denotes the responsivity of the \acp{PD} in A/W, which is assumed to be identical for all \acp{PD}. Furthermore, let $B$ and $N_0$ denote the bandwidth and the thermal noise power spectral density of a \ac{PD}, respectively, which are also assumed to be identical for all \acp{PD}\footnote{As noted later, the developed results and intuitions are also valid for unequal responsivities and noise power spectral densities.}. Moreover, we consider \ac{OWC} in the thermal noise limited regime where the impact of shot noise and relative intensity noise is considered negligible \cite{Sarbazi2024}. Hence, the \ac{SNR} for the $m$-th \ac{PD}, $m=1,\dots,M,$ is given by
\begin{align}
	\gamma_m = \frac{(R_{\mathrm{PD}} P^{\mathrm{Rx,O}}_m)^2}{B N_0}.
\end{align}

Next, we describe the area-bandwidth trade-off of \acp{PD}.

\subsection{Area-Bandwidth Trade-Off}
\label{sec:areabw}
The $3$-dB bandwidth of a \ac{PD} is given by $B = \frac{1}{\sqrt{(2\pi R_{\mathrm{L}} C_{\mathrm{j}})^2 + (2 \pi t_{\uptau})^2}}$ \cite{Alexander1997}, where $C_{\mathrm{j}} = \frac{\epsilon A}{d}$ denotes the junction capacitance and $\epsilon,$ $A,$ and $d$ denote the junction permittivity, area, and thickness, respectively, $R_{\mathrm{L}}$ denotes the load resistance, and $t_{\uptau}$ denotes the junction carrier transit time. As has been shown in \cite{Alexander1997}, when the bandwidth is limited by the junction capacitance, $B \propto \frac{1}{A}$ holds, while when it is limited by the carrier transit time, $B$ is independent of $A,$ and when the junction thickness is optimized for the maximum bandwidth, then $B \propto \frac{1}{\sqrt{A}},$ see also \cite{Sarbazi2024}. We consider all three regimes. Furthermore, in order to identify the most beneficial operating regime of the \ac{PD} array, we compare its performance with that of a single reference \ac{PD} of fixed bandwidth $B_0$ and the same total active area. This ensures that the effect of bandwidth expansion due to the changed active area of the \acp{PD} when scaling the \ac{PD} array is properly taken into account. Moreover, we assume that, in all three cases above, the initial bandwidth of the \ac{PD} array, i.e., for $M=1,$ is $B_0.$

Next, we define optical power distribution patterns, which will be considered throughout the letter.

\subsection{Considered Laser Beam Patterns}
\label{sec:degenerate}
\label{sec:lg}
\begin{figure}
	\centering
	\begin{minipage}{\ms{0.4\columnwidth}}
		\centering
		\includegraphics[width=\textwidth]{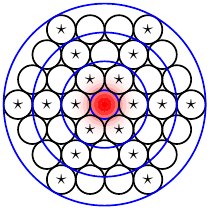}
	\end{minipage}%
	\begin{minipage}{\ms{0.4\columnwidth}}
		\centering
		\includegraphics[width=\textwidth]{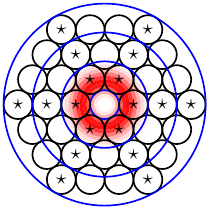}
	\end{minipage}
	\caption{Illustration of the beam power distribution for Gaussian and \acs{LG}$_{10}$ beams with $G=3$ rings and $\rho = 0.5$ on a \acs{PD} array employing the optimal hexagonal packing. \acp{PD} marked with a $\star$ denote the six corner \acsp{PD} in each ring.}
	\label{fig:gopd}
\end{figure}

In the following, we consider four laser beam patterns. The degenerate and uniform beam patterns serve as edge cases for the laser beam power distribution. On the other hand, Gaussian and \acs{LG}$_{10}$ beam patterns, which we consider as representative patterns for our study, are the more practical schemes.

For the degenerate beam pattern, we have $P^{\mathrm{Rx,O}}_1 = P^{\mathrm{Rx,O}}_{\mathrm{tot}}$ and $P^{\mathrm{Rx,O}}_m = 0,$ $m=2,\dots,M,$ up to permutation. That is, the \ac{PD} with index $1$ receives all the power and the other \acp{PD} do not receive any optical signal.

Next, for the uniform optical received power distribution, we have $P^{\mathrm{Rx,O}}_m = \frac{P^{\mathrm{Rx,O}}_{\mathrm{tot}}}{M},$  for $m=1,\dots,M.$ That is, all the \acp{PD} receive the same optical power.

Next, we consider the Gaussian\footnote{We note that the fundamental \ac{LG}$_{00}$ mode is equivalent to the Gaussian beam.} and \ac{LG}$_{10}$ beam patterns. We note that the Gaussian beam concentrates energy close to the center, whereas the \ac{LG}$_{10}$ beam has the energy concentrated in a ring around a central vortex, as shown in Figure \ref{fig:gopd}.

Furthermore, for these beams, since the captured energy depends on the placement of the \acp{PD} in the \ac{PD} array, we assume the optimal hexagonal packing\footnote{We note that, although in this letter we limit ourselves to representative \ac{LG} modes and the optimal hexagonal packing, the methods described can be extended to arbitrary modes and positions of \acp{PD} within the \ac{PD} array in a straightforward manner.}, which covers about $90\%$ of the circular area \cite{Fukshansky2011}. Furthermore, we assume that $M$ has the form $M = 1+3G(G+1),$ for some non-negative integer $G.$ This enables positioning a central \ac{PD} aligned to the incident beam center, and $G$ concentric hexagonal rings in the periphery, where the $g$-th ring contains $6g$ \acp{PD}. The pattern is illustrated in Figure \ref{fig:gopd} for the Gaussian and \ac{LG}$_{10}$ modes for $G=3$ and $\rho = 0.5.$ Throughout this letter, $\rho$ denotes the ratio of the \ac{PD} radius to the beam waist.

Next, based on \cite{Fukshansky2011,Azzolin2021}, for the Gaussian beam, the received optical power for the central \ac{PD} is given by
\begin{align}
	P^{\mathrm{Rx,O}}_1 = \frac{(1 - \mathrm{e}^{-2 \rho^2})}{1 - \mathrm{e}^{-2 (G+1)^2 \rho^2}} P^{\mathrm{Rx,O}}_{\mathrm{tot}},
\end{align}
where the expressions in the numerator and denominator are obtained by Gaussian radial integration and are the received optical powers for the central and reference \acp{PD}, respectively, and their ratio denotes the fraction of the total power received by the central \ac{PD}. Furthermore, the received optical power for the six corner \acp{PD} on ring $g,$ $g = 1,\dots,G,$ which are marked with a $\star$ in Figure \ref{fig:gopd}, is given by
\begin{align}
	P^{\mathrm{Rx,O}}_{\mathrm{Gauss,c},g} = \frac{(1 - Q_1(4 g \rho, 2 \rho))}{1 - \mathrm{e}^{-2 (G+1)^2 \rho^2}} P^{\mathrm{Rx,O}}_{\mathrm{tot}},
\end{align}
and for the remaining $(6g - 6)$ \acp{PD}, which are unmarked in Figure \ref{fig:gopd}, it is given by
\begin{align}
	P^{\mathrm{Rx,O}}_{\mathrm{Gauss,e},g} = \frac{(1 - Q_1(2\sqrt{3} g \rho, 2 \rho))}{1 - \mathrm{e}^{-2 (G+1)^2 \rho^2}} P^{\mathrm{Rx,O}}_{\mathrm{tot}}.
\end{align}
Here, $Q_m(a,b)$ denotes the Marcum's $Q$ function of order $m.$ Analogously, based on \cite{Kovalev2016,Simon2002}, for the \ac{LG}$_{10}$ beam, the received optical power for the central \ac{PD} is given by
\begin{align}
	P^{\mathrm{Rx,O}}_1 = \frac{(1 - \mathrm{e}^{-2 \rho^2} - 2\rho^2 \mathrm{e}^{-2 \rho^2})P^{\mathrm{Rx,O}}_{\mathrm{tot}}}{1 - \mathrm{e}^{-2 (G+1)^2\rho^2} - 2(G+1)^2\rho^2 \mathrm{e}^{-2 (G+1)^2 \rho^2}}.
\end{align}
Furthermore, the received optical power for the six corner \acp{PD} on ring $g,$ $g = 1,\dots,G,$ is given by
\begin{align}
	P^{\mathrm{Rx,O}}_{\mathrm{LG_{10},c},g} = \frac{\phi(2 g\rho, \rho)P^{\mathrm{Rx,O}}_{\mathrm{tot}}}{1 - \mathrm{e}^{-2 (G+1)^2\rho^2} - 2 (G+1)^2\rho^2 \mathrm{e}^{-2 (G+1)^2 \rho^2}},
\end{align}
and, for the remaining $(6g - 6)$ \acp{PD}, it is given by
\begin{align}
	P^{\mathrm{Rx,O}}_{\mathrm{LG_{10},e},g} = \frac{\phi(\sqrt{3}g\rho, \rho)P^{\mathrm{Rx,O}}_{\mathrm{tot}}}{1 - \mathrm{e}^{-2 (G+1)^2\rho^2} - 2(G+1)^2\rho^2 \mathrm{e}^{-2 (G+1)^2 \rho^2}},
\end{align}
where $\phi(a,b) = 2 a^2 (1 - Q_2(2 a, 2 b)) + (1 - Q_1(2 a, 2 b)) - 2 b^2 \mathrm{e}^{-2 (a^2 + b^2)} I_0(4 a b),$ and $I_0(x)$ is the modified Bessel function of the first kind and order zero.

Next, we derive the performance scaling laws.

\section{Combining in the Electrical Domain}
\label{sec:combining}

In this section, we analyze the impact of combining in the electrical domain and derive the performance scaling laws for \ac{IM/DD} systems employing \ac{PD} arrays.

\subsection{Optimal \acs{SNR}}
\label{sec:optimal}
First, we note that for maximal ratio combining, which is the optimal linear combining scheme, the \ac{SNR}, $\gamma_{\mathrm{mrc}},$ is given by $\gamma_{\mathrm{mrc}} = \sum_{m = 1}^{M} \gamma_m,$ whose maximum value depends on the input optical power distribution, which can be optimized as
\begin{argmaxi}
	{\footnotesize\makecell{\{P^{\mathrm{Rx,O}}_m\,,\forall\,m\}\\\sum_{m = 1}^{M} P^{\mathrm{Rx,O}}_m\leq P^{\mathrm{Rx,O}}_{\mathrm{tot}}}}
	{\gamma_{\mathrm{mrc}}.}
	{\label{opt:gs}}
	{\{P^{\mathrm{Rx,O}\star}_m\,,\forall\,m\}=}
\end{argmaxi}
Based on \cite[Theorem C.1 and Equation (8)]{Marshall2011}, Problem (\ref{opt:gs}) is Schur-convex, and the optimal received optical power distribution is degenerate\footnote{We note that, based on \cite[Theorem C.1 and Equation (8)]{Marshall2011}, the degenerate distribution is optimal also when the responsivity and noise variance of the \acp{PD} are not identical.}, up to permutation, see Section \ref{sec:degenerate}. Hence, for optimal performance, the received beam must be incident on a single \ac{PD} in the \ac{PD} array in order to minimize the loss due to the square-law relationship between the optical and electrical powers. In such a case, the optimal \ac{SNR} is given by $\gamma_{\mathrm{mrc}}^\star = \frac{(R_{\mathrm{PD}} P^{\mathrm{Rx,O}}_{\mathrm{tot}})^2}{B N_0}.$ Hence, beam patterns such as Gaussian, vortex beams, uniform beams, and other higher-order transverse electromagnetic beams do not achieve the maximum \ac{SNR} in general. Nevertheless, focusing all power on a single \ac{PD} of the \ac{PD} array is impractical due to alignment considerations and the law of \'etendue conservation, which disallows arbitrary concentration of beams. Next, in the following, we model the extent of \emph{suboptimality} of the considered practical beams.

\subsection{Loss Factor}
In order to derive insights for the practical case, where the received optical power distribution is not degenerate, we define the actually achieved (suboptimal) \ac{SNR} as
\begin{align}
	\gamma_{\mathrm{sub}} = \frac{(\beta R_{\mathrm{PD}} P^{\mathrm{Rx,O}}_{\mathrm{tot}})^2}{B N_0} = \beta^2 \gamma_{\mathrm{mrc}}^\star,
\end{align}
where $\beta,$ $0 \leq \beta \leq 1,$ is the \emph{loss factor,} which models the suboptimality of the \ac{PD} design and incident beam with a single parameter. In the following, we utilize the above model based on $\gamma_{\mathrm{sub}}$ for analyzing arbitrary beam power distributions, including Gaussian, \ac{LG}$_{10},$ and uniform beams, incident on the hexagonal \ac{PD} arrangement described earlier.

\subsection{Performance Comparison with the Reference \acs{PD}}
\label{sec:comparison}
From Section \ref{sec:areabw}, we note that the bandwidth $B$ of the \acp{PD} scales as $B = M^\xi B_0,$ where $\xi = 1, \frac{1}{2},$ and $0$ for capacitance limited, thickness optimized, and transit-time limited \acp{PD}, where the active area of the individual \acp{PD} is assumed to be $\frac{1}{M}$-th of the reference \ac{PD}. Hence, the achievable rate, $R,$ for the \ac{PD} array is given by the Shannon formula\footnote{We note that although we utilize the Shannon formula to analyze the impact of bandwidth on performance, the insights obtained in this paper are also applicable to the upper and lower bounds in \cite{Lapidoth2009,Chaaban2020}.} as
\begin{align}
	R = M^\xi B_0 \log\Big(1 + \beta^2\frac{\gamma_{\mathrm{mrc}}^\star}{M^\xi}\Big). \label{eqn:G}
\end{align}
For a \ac{PD} array with $M > 1$ \acp{PD} to have a better performance than the single reference \ac{PD},
\begin{equation}
	R \geq B_0 \log(1 + \gamma_{\mathrm{mrc}}^\star)
\end{equation}
must be satisfied, which leads to the condition
\begin{align}
	\beta^2 \geq \beta_{\mathrm{min}}^2 = \frac{M^\xi}{\gamma_{\mathrm{mrc}}^\star}\big[(1 + \gamma_{\mathrm{mrc}}^\star)^\frac{1}{M^\xi} - 1\big]. \label{eqn:condopt}
\end{align}
Hence, for $\beta^2 \geq \beta_{\mathrm{min}}^2,$ which is discussed in detail below, a \ac{PD} array meets or exceeds the performance of the single reference \ac{PD}. Therefore, ideally, $\beta_{\mathrm{min}}^2$ must be as low as possible to establish a low barrier. On the other hand, when $\beta^2 < \beta_{\mathrm{min}}^2,$ \ac{PD} arrays offer no performance benefit over the single reference \ac{PD}, but may have other advantages such as misalignment tolerance and fabrication ease, see, e.g., \cite{Koonen2020,Umezawa2021}, among others. 

\subsubsection{Impact of Scaling $\gamma_{\mathrm{mrc}}^\star$}
We note that $\lim_{\gamma_{\mathrm{mrc}}^\star \to \infty} \beta_{\mathrm{min}}^2 = 0$ for all $\xi.$ Hence, $\beta_{\mathrm{min}}^2$ can be made arbitrarily small by increasing the received \ac{SNR}, $\gamma_{\mathrm{mrc}}^\star,$ sufficiently. This enables us to meet or exceed the reference \ac{PD} performance for arbitrary $\beta^2,$ see Section \ref{sec:alpha}.

\subsubsection{Impact of Scaling $M$}
Next, we note that for $\xi = 1, \frac{1}{2},$ $\lim_{M \to \infty} \beta_{\mathrm{min}}^2 = \frac{\log(1 + \gamma_{\mathrm{mrc}}^\star)}{\gamma_{\mathrm{mrc}}^\star},$ indicating that for both capacitance and thickness optimized devices, $\beta_{\mathrm{min}}^2$ has an asymptotic floor. Hence, unlike with \ac{SNR}, $\beta_{\mathrm{min}}^2$ cannot be made arbitrarily small. Therefore, scaling $M$ alone is not sufficient to meet the reference \ac{PD} performance. On the other hand, for $\xi = 0,$ $\lim_{M \to \infty} \beta_{\mathrm{min}}^2 = 1,$ indicating that only the degenerate received optical power distribution meets the reference \ac{PD} performance.

\subsection{Loss Factor for the Considered Distributions}
\label{sec:betaprac}
For the degenerate distribution, $\beta_{\mathrm{degen}}^2 = 1,$ and for the uniform optical received power distribution, $\beta_{\mathrm{unif}}^2 = \frac{1}{M}.$ On the other hand, for a Gaussian beam, $\beta^2$ can be simplified to
\begin{align}
	\beta_{\mathrm{Gauss}}^2 &= \frac{(1-e^{-2\rho^{2}})^2 + \sum_{g=1}^{G} 6(1 - Q_1(4 g \rho, 2 \rho))^2}{(1 - \mathrm{e}^{-2 (G+1)^2 \rho^2})^2}\nonumber\\
	&\quad + \frac{\sum_{g=1}^{G}(6g-6)(1 - Q_1(2\sqrt{3} g \rho, 2 \rho))^2}{(1 - \mathrm{e}^{-2 (G+1)^2 \rho^2})^2},
\end{align}
and for the \ac{LG}$_{10}$ beam to
\begin{align}
	&\beta_{\mathrm{LG_{10}}}^2 = \frac{(1 - \mathrm{e}^{-2 \rho^2} - 2\rho^2 \mathrm{e}^{-2 \rho^2})^2 + \sum_{g=1}^{G} 6\phi(2 g\rho, \rho)^2}{(1 - \mathrm{e}^{-2 (G+1)^2\rho^2} - 2 (G+1)^2\rho^2 \mathrm{e}^{-2 (G+1)^2 \rho^2})^2}\nonumber\\
	&\quad\quad + \frac{\sum_{g=1}^{G}(6g-6)\phi(\sqrt{3}g\rho, \rho)^2}{(1 - \mathrm{e}^{-2 (G+1)^2\rho^2} - 2 (G+1)^2\rho^2 \mathrm{e}^{-2 (G+1)^2\rho^2})^2}.
\end{align}
We note that as $\rho \to 0$ and $G \to \infty,$  $\beta_{\mathrm{Gauss}}^2$ and $\beta_{\mathrm{LG_{10}}}^2$ approach $G^{-1}$ and $G^{-3},$ respectively. On the other hand, for $\rho > 1,$ the whole beam waist falls within the central \ac{PD}, resembling the degenerate beam, see also Figure \ref{fig:gopd}. Hence, $\rho \to 0$ represents the practically relevant case.

\subsubsection{Central \acs{PD} Only}
For the case where only the central \ac{PD} is utilized, whose bandwidth, unlike the reference \ac{PD}, scales with the active area, then, as $\rho \to 0,$ $\beta_{\mathrm{Gauss},1}^2$ and $\beta_{\mathrm{LG_{10}},1}^2$ approach $G^{-2}$ and $G^{-4},$ respectively. Hence, they are asymptotically inferior compared to utilizing the entire \ac{PD} array\footnote{We note that the ``Central \acs{PD} Only'' case is impractical for \acs{LG}$_{10}$ since the central \acs{PD} always lies in the vortex.}.

\subsubsection{Higher-Order \acs{LG} Modes}
We note that $\frac{\beta_{\mathrm{Gauss}}^2}{\beta_{\mathrm{LG_{10}}}^2} \sim G^2$ as $\rho \to 0, G\to\infty,$ which shows that the Gaussian beam is asymptotically superior compared to the \ac{LG}$_{10}$ beam owing to the vortex of the latter and the greater spreading of the incident power, see also Section \ref{sec:numerical}. The same insight applies to other higher order modes, which spread their incident power over larger areas compared to the Gaussian beam.

\subsubsection{Received Power Scaling}
\label{sec:alpha}
For a design with loss factor $\beta^2,$ the performance of the reference \ac{PD} can be met by increasing the received \emph{optical} power by $\alpha = \sqrt{\frac{\beta_{\mathrm{min}}^2}{\beta^2}}.$ We note that for the practically relevant case of $\rho \to 0,$ we have $\lim_{\rho\to 0}\lim_{G\to\infty} \alpha_{\mathrm{Gauss}} = \infty$ and $\lim_{\rho\to 0}\lim_{M\to\infty} \alpha_{\mathrm{LG_{10}}} = \infty,$ i.e., for suboptimal systems with $\beta^2 < \beta_{\mathrm{min}}^2,$ the received power must continually increase with increasing $G$ in order to meet the reference \ac{PD} performance.

\section{Numerical Results and Discussion}
\label{sec:numerical}
In this section, we present numerical results, along with a discussion.

\begin{figure}
	\centering
	\includegraphics[width=\ms{0.85\columnwidth}]{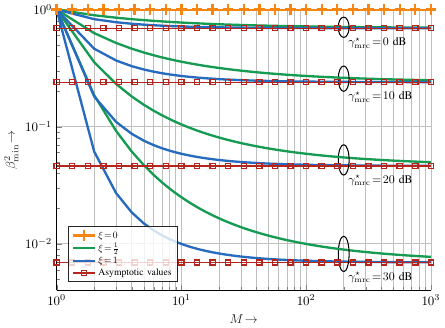}
	\caption{Minimum required loss factor $\beta_{\mathrm{min}}^2$ as a function of $M$ for different \acsp{SNR}, based on (\ref{eqn:condopt}).}
	\label{fig:betamin}
\end{figure}

First, we study the minimum loss factor $\beta_{\mathrm{min}}^2,$ which incorporates the effects of bandwidth expansion and \ac{SNR} via $\xi$ and $\gamma_{\mathrm{mrc}}^\star,$ respectively, as a function of the number of \acp{PD} $M.$ Figure \ref{fig:betamin} illustrates $\beta_{\mathrm{min}}^2$ as a function of $M$ for different \ac{SNR} and $\xi$ values, based on (\ref{eqn:condopt}). The figure shows that, for $\xi = 1, \frac{1}{2},$ $\beta_{\mathrm{min}}^2$ converges rapidly to the respective asymptotic value given in Section \ref{sec:comparison}, indicating that increasing $M$ alone yields no improvement over the single reference \ac{PD} performance. Furthermore, for small $M,$ thickness-optimized \acp{PD} are slightly inferior compared to capacitance limited ones. Nevertheless, they perform similarly asymptotically. Moreover, as the \ac{SNR} increases, the required $\beta_{\mathrm{min}}^2$ decreases significantly. This validates the analytical observation that \ac{PD} arrays become beneficial above a certain \ac{SNR} threshold, where the bandwidth gain achieved by reducing the \ac{PD} area outweighs the electrical power loss caused by the square-law relationship, see Section \ref{sec:combining}. On the other hand, for $\xi = 0,$ $\beta_{\mathrm{min}}^2$ is independent of $M.$

\begin{figure}
	\centering
	\includegraphics[width=\ms{0.85\columnwidth}]{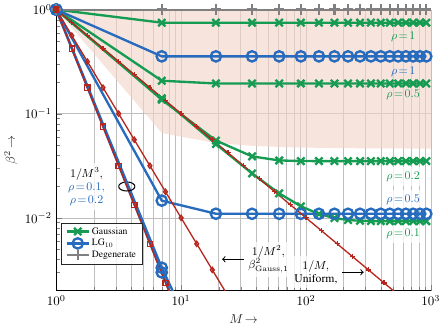}
	\caption{Loss factor $\beta^2$ versus the number of photodetectors $M$ for Gaussian and \ac{LG}$_{10}$ beams with fixed PD-radius-to-beam-waist ratio $\rho.$}
	\label{fig:betalg}
	\vspace{-0.25cm}
\end{figure}

Next, in the following figures, we study the the loss factor $\beta^2,$ which models the effects of beam shape and parameters, array \ac{PD} positions, and the impact of square-law. Figure \ref{fig:betalg} shows $\beta^2$ versus $M$ for Gaussian and \ac{LG}$_{10}$ beams\footnote{Only values of $M = 1 + 3G(G+1)$ are reported.} when the PD-radius-to-beam-waist ratio $\rho$ is kept constant, along with the asymptotic values discussed in Section \ref{sec:combining}. In this case, with increasing $M,$ the overall active area of the \ac{PD} array increases. The results show that for small $\rho$, both modes exhibit a rapid decay of $\beta^2,$ with \ac{LG}$_{10}$ degrading substantially faster than Gaussian owing to its vortex and greater power spread. Comparatively, the uniform beam experiences an intermediate performance degradation. For larger $\rho$, where most of the optical power is concentrated on the central \ac{PD}, $\beta^2$ approaches unity, which is consistent with the degenerate distribution. These results highlight the strong dependence of performance on beam concentration and validate the suboptimality of direct detection of higher-order modes. Lastly, in this and the next figures, pink shading denotes the region in which $\beta^2 \geq \beta_{\mathrm{min}}^2$ for $\xi = 1$ and  $\gamma_{\mathrm{mrc}}^\star = 20$~dB.

\begin{figure}
	\centering
	\includegraphics[width=\ms{0.85\columnwidth}]{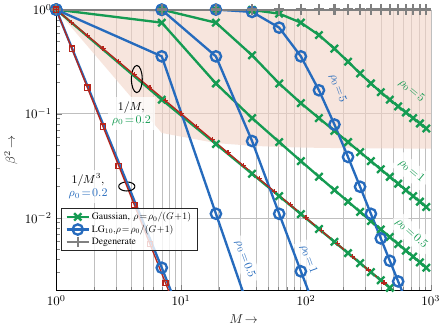}
	\caption{Loss factor $\beta^2$ versus the number of photodetectors $M$ when the \acs{PD} radius scales inversely with array size, $\rho=\rho_0/(G+1).$}
	\label{fig:betalgrho}
	\vspace{-0.25cm}
\end{figure}

Next, Figure \ref{fig:betalgrho} considers a more practical scenario, where $\rho$ decreases with increasing array size as $\rho=\rho_0/(G+1).$ This corresponds to the case where the \ac{PD} radius shrinks with $M,$ while the reference \ac{PD} size remains constant. Here, the effective beam footprint becomes increasingly spread over the \ac{PD} array as $M$ grows. Hence, $\beta^2$ degrades for increasing $M$ for both  modes, and the gap between Gaussian and \ac{LG}$_{10}$ widens. This result shows that, for realistic receiver scaling, increasing the number of \acp{PD} exacerbates the square-law penalty unless the received beam remains sufficiently narrow.

Nevertheless, in both cases, the performance of the reference \ac{PD} is exceeded whenever $\beta^2 > \beta_{\mathrm{min}}^2$ for the designed system parameters. Furthermore, the results are consistent with the analysis in Section \ref{sec:combining}. Gaussian beams behave like uniform beams for small $\rho$ and like degenerate beams for large $\rho$, whereas higher-order LG modes have a substantially larger penalty due to their spatially distributed intensity profiles. Therefore, for \ac{PD} array-based \ac{IM/DD} receivers, meeting or exceeding the performance of a single reference \ac{PD} generally requires careful parameter choice or increasing the received power rather than merely scaling the number of \acp{PD}. Furthermore, based on the large performance penalty, we note that converting higher-order spatial modes to the fundamental Gaussian mode prior to direct detection is advantageous in practice. The above results highlight the need for joint optimization of beam pattern, transverse electromagnetic mode, received power, and PD positions.

\section{Conclusion}
\label{sec:conclusion}
We studied performance scaling laws for PD array-based receivers in \ac{IM/DD} optical wireless systems by explicitly accounting for the inherent square-law relationship between optical and electrical powers. By comparing \ac{PD} arrays to a reference single-\ac{PD} receiver for Gaussian, \ac{LG}$_{10},$ and uniform beam patterns and for different \ac{PD} area scaling assumptions, we showed that \ac{PD} arrays provide performance gains for sufficiently narrow beams and above an \ac{SNR} threshold, depending on the loss factor $\beta^2.$ Furthermore, increasing $M$ alone does not exceed reference \ac{PD} performance, and, in most cases, received optical power must be scaled by $\sqrt{\frac{\beta_{\mathrm{min}}^2}{\beta^2}}$ to meet it. Moreover, direct detection of higher-order modes, which spread power over several \acp{PD}, is generally suboptimal, and conversion to the fundamental Gaussian beam before detection is beneficial. These results provide clear guidelines and highlight the trade-offs in designing next-generation high-bandwidth PD array receivers.

\bibliographystyle{IEEEtran}
\bibliography{references}

\end{document}